\documentclass[aps,pra,twocolumn,superscriptaddress,floatfix]{revtex4-2}
\usepackage[T1]{fontenc}
\usepackage{graphicx}
\usepackage{color}
\usepackage{bm}
\usepackage{longtable}
\usepackage{comment}
\usepackage{amsmath}
\usepackage{amsfonts}
\usepackage{amssymb}
\usepackage{subfigure}
\usepackage{bbold}
\usepackage{bbm}
\usepackage{rotating}
\usepackage{verbatim}
\usepackage{braket}
\usepackage{multirow}
\usepackage{hyperref}
\usepackage[english]{babel}
\usepackage{comment}
\usepackage{float}
\usepackage{ulem}
\usepackage{blkarray, bigstrut}
\usepackage{color}

\begin{document}

\title{Dynamical discontinuities in repeated weak measurements revealed by complex weak values}

\author{Lorena Ballesteros Ferraz}
\affiliation{Laboratoire de Physique Théorique et Modélisation (LPTM) – UMR CNRS 8089, CY Cergy Paris Université, Cergy-Pontoise, France}
\date{\today}
\email{lorena.ballesteros-ferraz@cyu.fr}

\begin{abstract}
This work demonstrates that repeated weak measurements together with post-selection can produce sharp dynamical discontinuities in meter observables, even in minimal quantum systems. The discontinuous behavior is governed by the polar angle of the post selected state, which serves as a continuous control parameter. As this angle is varied, the expectation value of the meter observables changes abruptly at the point where the imaginary part of the associated weak value, a complex quantity that arises in weak measurements with post-selection, becomes zero. Such non-analytic behavior emerges only when the weak value is genuinely complex for a range of post-selection angles. If the weak value remains purely real for all angles, the dynamics remain smooth. The discontinuity originates from an exchange of stability between fixed points of the non-unitary Kraus operator governing the meter’s evolution. Remarkably, despite the absence of a thermodynamic limit, the relaxation time in the vicinity of the discontinuity exhibit universal critical behavior characterized by a critical exponent equal to $1$, independent of system parameters. These results establish the weak value as a tunable control parameter capable of inducing non-analytic dynamical responses and reshaping the stability structure of measurement-induced quantum dynamics.
\end{abstract}

\maketitle

\section{Introduction}
Weak measurements with post-selection, introduced by Aharonov and colleagues~\cite{aharonov1988result}, consist of three steps. First, the system is prepared in a known initial state, a procedure referred to as pre-selection. Second, it undergoes a weak and short interaction with a measurement device, described by a unitary operator with a small coupling strength. Finally, a specific final state of the system is post-selected through a projective measurement followed by filtering. In this setting, the meter shift is proportional to the weak value of the observable, a complex quantity that can be anomalous, meaning it may lie outside the range of the observable’s eigenvalues. For example, it can take values larger or smaller than the eigenvalue spectrum or have a nonzero imaginary part.

The physical interpretation of complex weak values has been explored in the literature. Jozsa analyzed how both the real and imaginary parts of the weak value manifest in measurable shifts of the pointer’s position and momentum in a weak measurement context~\cite{jozsa2007complex}. Hofmann showed that the complex phases that arise in weak conditional probabilities describe the dynamical response of a system to small unitary transformations generated by the measured observable~\cite{hofmann2011role, hofmann2012complex}. These studies connect the complex nature of weak values with concrete measurement outcomes and underlying dynamical properties of quantum systems.

Anomalous weak values, whose values are outside the observable spectrum, have proven useful both in foundational studies of quantum theory~\cite{aharonov2002revisiting, matzkin2012observing, dressel2015weak, ferraz2022geometrical, ferraz2023revisiting, ferraz2025exploring} and in enhancing sensitivity to weak signals in precision measurement and sensing applications~\cite{jordan2014technical, dixon2009ultrasensitive, luo2017precision, ferraz2024relevance}. Furthermore, weak values are closely connected to the Kirkwood–Dirac quasiprobability distribution, which exhibits nonclassical features whenever weak values are anomalous. Anomalous weak values have been associated with quantum contextuality, as well as with counterfactual and coherence effects in pre- and post-selected scenarios~\cite{arvidsson2024properties, pusey2014anomalous, kunjwal2019anomalous, hance2024counterfactuality, hance2023contextuality}.

The study of non-equilibrium quantum dynamics has uncovered a rich variety of critical phenomena that extend the concept of phase transitions outside the realm of equilibrium physics. While traditional equilibrium transitions are driven by temperature or external fields, such as magnetic fields~\cite{sachdev1999quantum, Hertz1976, millis1993effect, vojta2003quantum}, non-equilibrium quantum systems can exhibit purely quantum phase transitions with no classical analogue. These transitions may manifest in the dynamics of time-evolving observables~\cite{heyl2013dynamical, vzunkovivc2016dynamical} or in measures of quantum entanglement~\cite{de2021entanglement, kawabata2023entanglement}.

Dynamical phase transitions arise in quantum systems driven out of equilibrium, typically after a sudden change in a system parameter, called a quantum quench. Unlike conventional phase transitions, which involve abrupt changes in thermodynamic quantities as external control parameters are varied, dynamical phase transitions are marked by non-analytic behavior in time-dependent observables. A common example is the Loschmidt echo, which quantifies the overlap between the initial state and its time-evolved state; sharp kinks or sudden changes in this quantity indicate a qualitative change in the system’s dynamics. Such transitions have been observed in spin chains, cold-atom systems, and trapped-ion simulators~\cite{heyl2013dynamical, zvyagin2016dynamical, jurcevic2017direct}.

Measurement-induced phase transitions emerge from the competition between unitary entanglement generation and measurement-induced decoherence. While unitary dynamics tend to spread entanglement, repeated measurements suppress it, leading to a phase transition between highly entangled and weakly entangled states with universal scaling behavior~\cite{skinner2019measurement, block2022measurement, li2023cross}. These transitions have been observed in both numerical simulations and quantum experiments~\cite{noel2022measurement, koh2023measurement}, and in many-body systems they exhibit genuine non-analytic behavior, most prominently in entanglement measures, as the measurement rate or strength is varied~\cite{Szyniszewski2019}. 

Remarkably, signatures of criticality and non-analytic behavior are not restricted to the thermodynamic limit. Even small systems, such as two-level or Rabi-type models, can display sudden dynamical changes when system parameters are tuned across critical regimes. This indicates that essential features of quantum criticality, including abrupt transitions in observables and entanglement, persist in few-body settings~\cite{puebla2020finite, hamazaki2021exceptional, di2024environment}.

The study of quantum dynamics under repeated measurements has a long history, ranging from rigorous analysis of quantum non-demolition protocols~\cite{Bauer2011,Bauer2012} to the investigation of measurement-induced phase transitions in many-body systems~\cite{Szyniszewski2019,Dubey2021}. In quantum non-demolition and weak-measurement settings, a system repeatedly interacts with probe degrees of freedom, and each measurement updates the state according to a completely positive map (a Kraus map) describing the measurement back-action. Iterating this map typically drives the system toward a stable fixed point or steady state, with the rate of convergence determined by the eigenvalues of the map. Importantly, in such scenarios the dependence on control parameters, such as measurement strength, is generally smooth, leading to gradual relaxation rather than abrupt changes in observable expectation values. More recently, repeated weak measurements of non-conserved observables have been shown to induce partial state collapse and emergent steady-state structures~\cite{Roux2024}, although the evolution of observables remains continuous as system parameters are tuned.

This work investigates the dynamics of quantum systems under repeated weak measurements with post-selection, where sharp discontinuities can appear in the expectation values of meter observables. The protocol shares certain features with measurement-induced and dynamical phase transitions, yet also differs in important ways. Unlike these phase-transition protocols, the focus here is on small systems. In contrast to measurement-induced transitions, which are usually associated with entanglement dynamics, this study focuses on expectation values, despite the presence of measurements. There is also an analogy with dynamical phase transitions, where post-selection plays a role similar to a quench, although the control parameter is not time. While weak measurements with post-selection have been linked to mathematical features reminiscent of quantum criticality, such as non-normality~\cite{ferraz2023revisiting}, they have not previously been connected to critical phenomena, nor has the weak value been systematically studied as an indicator of such behavior.

The protocol proceeds as follows. In each iteration, the system is initialized in a chosen state (pre-selection) and interacts weakly with a meter via a short, low-strength unitary coupling. Afterward, the system is post-selected in a desired state and discarded, while the meter is retained to interact with a freshly initialized system. Repeating this procedure $N$ times allows evaluation of the expectation value of a meter observable, for example, a Pauli operator $\hat{\sigma}$ in a two-level system, as a function of the post-selection polar angle. Discontinuities arise precisely when the imaginary part of the weak value vanishes, provided that (i) the weak value is not purely real for all post-selection angles, and (ii) the Kraus operator governing the meter evolution has non-degenerate eigenvalues.

The discontinuity is triggered when the imaginary component of the weak value crosses zero, inducing an exchange-of-stability bifurcation in the Kraus map: two dynamical fixed points interchange their stability, producing an abrupt change in the long-time behavior of the meter observable.

Extending the analogy with phase transitions, the relaxation dynamics near the discontinuity exhibit power-law scaling with a critical exponent of 1. Specifically, as the control parameter, the post-selection polar angle, approaches its critical value where the imaginary part of the weak value vanishes, the characteristic relaxation time, defined as the number of measurement iterations needed for the meter observable to reach its steady-state value, diverges inversely with the distance from the critical point. The fact that this exponent is the same for all system parameters indicates a form of universality. The scaling behavior depends only on the structure of the underlying dynamical map and not on details such as the choice of initial state.

Repeated weak measurements with post-selection can produce abrupt changes in long-time meter observables, revealing weak values as tunable control parameters and enabling high-sensitivity detection, controlled manipulation of the meter state, and the exploration of critical-like dynamics even in small quantum systems.

This paper is organized as follows: section~\ref{section:presentation_of_the_system_general} introduces the system and analyzes the Kraus operator associated with the protocol. Section~\ref{section:fixed_points_analysis} examines the fixed points of the protocol and their stability. Section~\ref{section:two_level_system} specializes the discussion to the two-level case. Section~\ref{section:analysis_discontinuity} presents the analysis of the critical exponent associated with the observed discontinuities, and Section~\ref{section:conclusion} summarizes the main conclusions.
\section{Protocol presentation}\label{section:presentation_of_the_system_general}
This section presents the protocol that generates discontinuities and stability bifurcations and analyzes the corresponding Kraus operator. Several qubits serve as the “systems” in a standard weak measurement protocol with post-selection, while an $N$-level system acts as the meter, the system in which discontinuities are induced.

The system starts in the state $\ket{\psi_S}$, while the meter is initialized in the state $\ket{\phi_A}$. The joint initial state is
\begin{equation}
    \ket{\Psi} = \ket{\psi_S} \otimes \ket{\phi_A}.
\end{equation}
The system and meter interact weakly through a unitary operator $\hat{U}$ which, assuming the interaction strength $g$ and the time $t$ are sufficiently small, can be expanded to first order in $gt$ as
\begin{equation}
    \hat{U} = e^{-igt \hat{O}_S \otimes \hat{O}_A} \approx \hat{I}\otimes\hat{I} - igt \, \hat{O}_S \otimes \hat{O}_A,
\end{equation}
where $\hat{O}_S$ acts on the system and $\hat{O}_A$ on the meter.

After the interaction, the system is post-selected onto the state $\ket{\psi_f}$. The meter state $\hat{\rho}_A^f$ after applying once the weak measurement protocol is obtained via the application of the Kraus operator $\hat{K}$ to the initial meter state:
\begin{equation}
   \hat{\rho}_A^{f,1} = \frac{\hat{K} \ket{\phi_A}\bra{\phi_A} \hat{K}^\dagger}{\text{Tr}\!\left[\hat{K} \ket{\phi_A}\bra{\phi_A} \hat{K}^\dagger \right]},
\end{equation}
where $\hat{K}$ is, in general, non-unitary and, to first order in $g$, can be expressed as
\begin{equation}
\label{eq:Kraus_operator}
    \hat K=\braket{\psi_f|\psi_S}\left[\hat{I}-igtO_{S,w}\hat{O}_A\right], 
\end{equation}
where
\begin{equation}
    O_{S,w}=\frac{\bra{\psi_f}\hat{O}_S\ket{\psi_S}}{\braket{\psi_f|\psi_S}}
\end{equation}
is the weak value of the observable $\hat{O}_S$. Accurately modeling the weak interaction and post-selection with the Kraus operator in Eq.~\ref{eq:Kraus_operator} demands more than the condition $gt \ll 1$. It also requires that the following condition is satisfied~\cite{kofman2012nonperturbative}
\begin{equation}
|gt\, O_{S,w}|\left( |\langle \hat{O}_A \rangle| + \Delta \hat{O}_A \right) \ll 1,
\end{equation}
where $\langle \hat{O}_A \rangle$ and $\Delta \hat{O}_A$ denote the expectation value and standard deviation of the ancilla observable $\hat{O}_A$ in its initial state.

\begin{figure} [h!]
    \centering
\includegraphics[width=0.5
\linewidth]{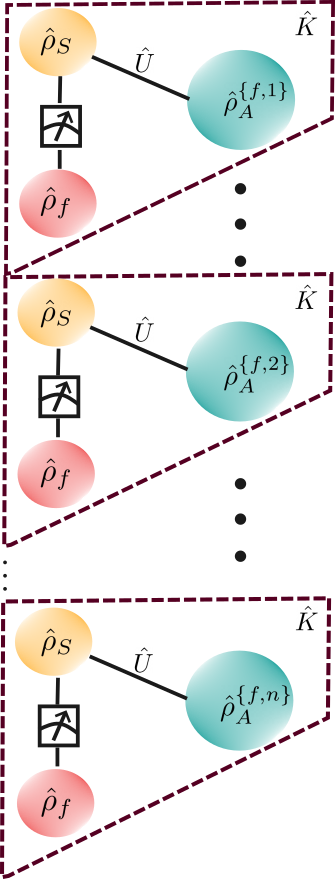}
    \caption{The protocol consists of the application of $n$ times the weak measurement protocol where the system is pre-selected $(\hat\rho_S)$, then it interacts with the meter (in the state $\hat\rho_A^{f,i}$) via a weak unitary operator and finally the system is post-selected ($\hat\rho_f$). After this protocol a new system is pre-selected ($\hat\rho_S$) and the protocol continues using the same meter.}\label{fig:protocol_drawing}
\end{figure}

The Kraus operator remains non-unitary at first order in $g$ as long as the imaginary part of the weak value is nonzero. To generate a dynamical process and observe discontinuities in the expectation value of certain observables in the meter’s final state, the weak measurement protocol is applied $n$ times, each involving a different system prepared in the same initial state, as illustrated in Fig.~\ref{fig:protocol_drawing}. Consequently, the meter state after the $n$ applications is
\begin{equation}
\hat{\rho}_A^{f,n}=\frac{\hat{K}^{n}\ket{\phi_A}\bra{\phi_A} \hat{K}^{\dagger ! n}}{\text{Tr}\!\left[\hat{K}^{n}\ket{\phi_A}\bra{\phi_A} \hat{K}^{\dagger ! n}\right]},
\end{equation}
where the Kraus operator $\hat{K}$ has been applied $n$ times to the initial meter state.

The weak measurement protocol to reach the discontinuity (Fig.~\ref{fig:protocol_drawing}), consisting of pre-selection, a weak unitary interaction, and post-selection, is applied to the meter, with the entire process represented by the Kraus operator $\hat K$. After each full application of the weak measurement protocol, the system is discarded and replaced with a new one for the next iteration (or reset to the initial state), while the meter is preserved throughout all $n$ applications.
The eigenvalues of the Kraus operator $\hat{K}$ are
\begin{equation}
\label{eq:general_eigenvalues_Kraus_operator}
    \lambda_{j}=\braket{\psi_f|\psi_S}\left[1-igtO_{S,w}o_{A, j}\right],
\end{equation}
where $j$ runs from 1 to $N$, with $N$ denoting the dimension of the operator, and $o_{A,j}$ representing the eigenvalues of the $N$-level operator $\hat{O}_A$. To the first order in $gt$, their moduli are
\begin{equation}
    |\lambda_{j}|=|\braket{\psi_f|\psi_S}|\sqrt{1-2gt\;\text{Im}{\left(O_{S,w}\right)}\;o_{A, j}}.
\end{equation}
Thus, the moduli of the eigenvalues are equal, $ |\lambda_{j}| = |\lambda_{j'}| $, whenever $\text{Im}(O_{S,w}) = 0$, i.e., when the imaginary part of the weak value vanishes, or when the operator $\hat{O}_A$ is degenerate. If $ \hat{O}_A $ is non-degenerate, a transition from non-unitary to unitary dynamics at first order in $gt$ occurs at $\text{Im}(O_{S,w}) = 0 $. At the point where the absolute values of the eigenvalues become equal, discontinuities may appear in the expectation values of observables in the meter state after applying the weak measurement protocol $n$ times. Such discontinuities emerge precisely at the points where the absolute values of the eigenvalues coincide (Appendix~\ref{appendix:analysis_discontinuities}).

This behavior emphasizes the essential role of complex weak values: discontinuities in the dynamics can arise only when $ O_{S,w} $ possesses a nonzero imaginary component for some post-selected polar angle. Conversely, if the weak value is purely real, all $N$ eigenvalues coincide for every value of $\phi$ (the post-selected polar angle), and no discontinuities occur since the eigenvalues remain equal.
\section{Fixed points and their stability}\label{section:fixed_points_analysis}
Let us consider the Kraus operator $\hat{K}$ associated with the weak measurement protocol with post-selection applied to the meter state, as defined in Eq.~\ref{eq:Kraus_operator}. Under the assumption that the initial meter state is pure, the Kraus map ensures that the system evolves only within the manifold of pure states, which, for a two-level system, is geometrically represented by the surface of the Bloch sphere. When the dynamics consist of repeated applications of $\hat{K}$, the fixed points of the evolution, those states that remain invariant once reached, correspond to the eigenvectors of $\hat{K}$. These eigenvectors coincide with those of the observable $\hat{O}_A $, denoted by $|a_j\rangle$, and defined by
\begin{equation}
\hat O_A |a_j\rangle = o_{A,j} |a_j\rangle, 
\qquad \lambda_j \in \mathbb{R},
\end{equation}
where $j$ runs from 1 to $N$, the dimension of the Hilbert space of the operator $\hat O_A$, and $o_{A,j}$ are its eigenvalues.

To analyze the stability of the process, $n$ successive applications of the Kraus operator $\hat{K}$ are considered. The initial meter state can be expressed in the eigenbasis of the observable $\hat{O}_A$ as  
\begin{equation}
\ket{\phi_A} = \sum_{j} c^{(0)}_{j} \ket{a_j},
\end{equation}
where $c^{(0)}_{j}$ denotes the initial amplitude associated with the eigenstate $\ket{a_j}$.

After $n$ iterations of the weak measurement protocol, corresponding to $n$ normalized applications of the Kraus operator $\hat{K}$, the probability of finding the meter in the component state $\ket{a_j}$ evolves as  
\begin{equation}
\label{eq:meter_probability_evolution}
|c_{j}^{(n)}|^2 =
\frac{|c_{j}^{(0)}|^2 \, \left| 1 + 2 g t \, \text{Im}(O_{S,w}) \, o_{A,j} \right|^{2n}}
{\sum_k |c_{k}^{(0)}|^2 \left| 1 + 2 g t \, \text{Im}(O_{S,w}) \, o_{A,k} \right|^{2n}},
\end{equation}
where $o_{A,j}$ is the eigenvalue of $\hat{O}_A$ associated with $\ket{a_j}$, and $O_{S,w}$ denotes the weak value of the system observable.

The meter state after $n$ applications is therefore  
\begin{equation}
\ket{\phi_A^{(n)}} = \sum_j c_j^{(n)} \ket{a_j},
\end{equation}
with coefficients $c_j^{(n)}$ resulting from $n$ successive normalized applications of the Kraus operator $\hat{K}$, the amplitudes $|c_j^{(n)}|$ of which are determined by Eq.~\ref{eq:meter_probability_evolution}.

As \( n \to \infty \), the system approaches a long-time limit characterized by a stable fixed point, which is the state toward which all other states converge, except for those corresponding to unstable fixed points. This stable state corresponds to the eigenvector of the Kraus operator associated with the eigenvalue of largest modulus, provided that the spectrum of the Kraus operator is non-degenerate. The remaining eigenvectors correspond to unstable fixed points.

Because \( g t \) is always positive, the sign of the imaginary part of the weak value $\text{Im}\left(O_{S,w}\right)$ determines which eigenvector becomes stable. Specifically:
\begin{itemize}
    \item If \( \text{Im}(O_{S,w}) < 0 \), the stable state is the eigenvector with the largest eigenvalue \( o_{A,+} \).
    \item If \( \text{Im}(O_{S,w}) > 0 \), the stable state is the eigenvector with the smallest eigenvalue \( o_{A,-}\).
\end{itemize}
Consequently, a discontinuity in the dynamics arises when the stability of the fixed points changes. This occurs precisely when the imaginary part of the weak value $\text{Im}(O_{S,w})$ crosses zero, defining a critical post-selected polar angle \( \phi_c \).

When the weak value has no imaginary part, $\text{Im}(O_{S,w})=0$, the situation changes qualitatively. In this case, no eigenvalue is favored over the others, and the evolution becomes effectively unitary at first order in $gt$, leading only to phase accumulation. 
\section{Qubit-qubit example}\label{section:two_level_system}
Consider a concrete example where both the meter and the system are qubits (two-level systems), following the same weak measurement setup as before. The initial system state lies on the meridian with zero phase, as given in the equation below.
\begin{equation}
    \ket{\psi_S(\theta)} = \cos\theta \ket{0} + \sin\theta \ket{1},
\end{equation}
whereas the post-selected state is an arbitrary qubit state
\begin{equation}
    \ket{\psi_f(\phi,\alpha)} = \cos\phi \ket{0} + e^{i\alpha}\sin\phi \ket{1}.
\end{equation}
The unitary interaction is defined as
\begin{equation}
\label{eq:unitary_operator}
\hat{U}(\gamma) = e^{-i\gamma t \hat{\sigma}_z \otimes \hat{\sigma}_x}=\cos{\gamma t} \left(\hat{I}\otimes\hat{I}\right)-i\sin{\gamma t}\left(\hat{\sigma}_z\otimes\hat{\sigma}_x\right)
\end{equation}
that is valid for all interaction strengths. The Kraus operator becomes
\begin{equation}
\label{eq:Kraus_general_strength}
    \hat{K} = c(\gamma)\hat{I} + d(\gamma)\hat{\sigma}_x,
\end{equation}
with
\begin{eqnarray}
    c(\gamma) &=& \cos{\gamma t}\;\braket{\psi_f|\psi_i} \\ \nonumber
		&=& \cos{\gamma t}\; \left(\cos\theta\cos\phi + e^{-i\alpha}\sin\theta\sin\phi\right) \\ \nonumber
    d(\gamma) &=& -i \sin{\gamma t}\; \braket{\psi_f|\psi_i} \sigma_{z,w} \\ \nonumber
		&=&	-i \sin{\gamma t} \;\left(e^{-i\alpha}\sin\theta\sin\phi - \cos\theta\cos\phi\right).
\end{eqnarray}
The corresponding eigenvalues are
\begin{equation}
\label{eq:eigenvalues_K}
    \lambda_{\pm} = c(\gamma) \pm d(\gamma).
\end{equation}
\begin{figure} \centering \includegraphics[width=0.9\linewidth]{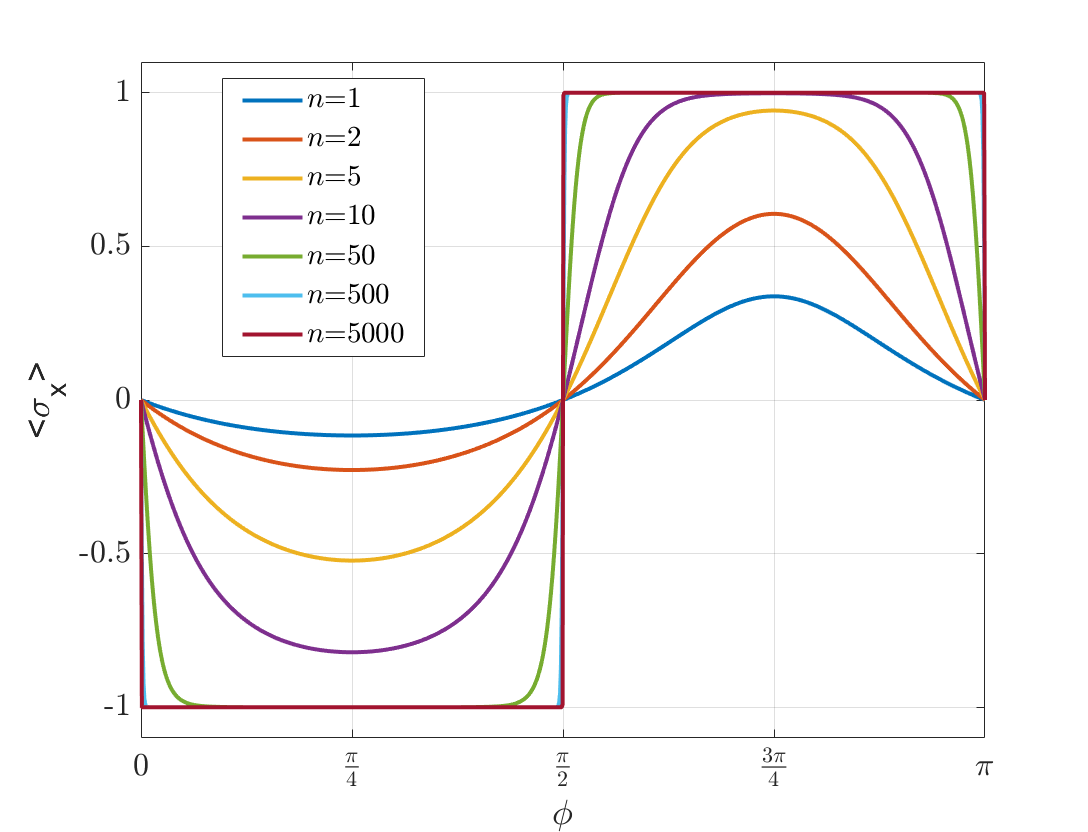} \caption{
Three distinct discontinuities characterize the dependence of the expectation value of $\hat\sigma_x$ on the post-selected angle $\phi$, emphasizing the nontrivial structure that arises for different repetition numbers $n$. In this plot $\alpha=\frac{\pi}{3}$, $\theta=\frac{\pi}{4}$ and $\gamma=0.1$, while the initial meter state is along the $z$ axis.} \label{fig:expectation_value_X} \end{figure}
For a single repetition, the expectation value of the observable $\hat{\sigma}_x$ for the final meter state
displays a smooth oscillatory behavior as a function of the post-selected polar angle (Fig.~\ref{fig:expectation_value_X}). As the number of repetitions of the protocol increases, however, the extrema become sharper, eventually forming a profile characterized by three distinct discontinuities. These discontinuities occur at the values of $\phi$ where the absolute values of the eigenvalues coincide, corresponding to the points where the imaginary part of the weak value vanishes, namely $\phi = 0$, $\phi = \frac{\pi}{2} $, and $\phi = \pi $. 
\begin{figure}
    \centering
\includegraphics[width=0.9\linewidth]{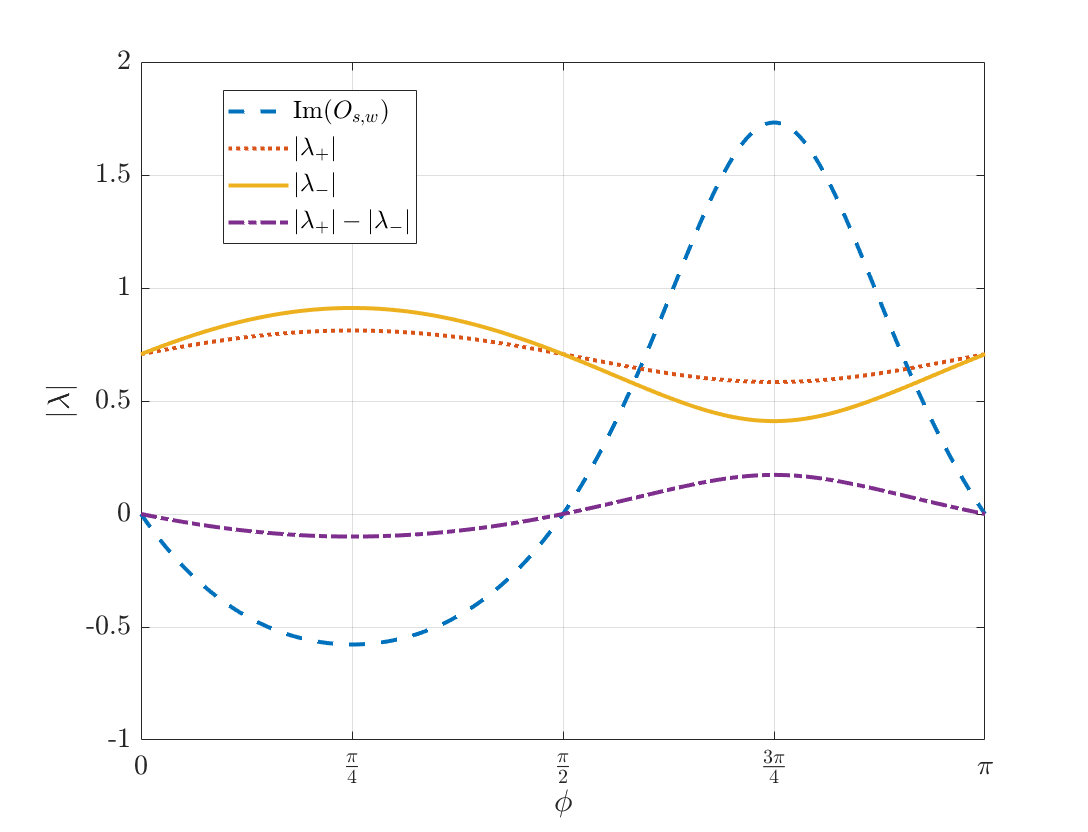}
    \caption{The imaginary part of the weak value vanishes when the absolute value of the two eigenvalues of the Kraus operator coincide. The plot shows the eigenvalues of the Kraus operator $\hat K$, $|\lambda_{+}|$ and $|\lambda_{-}|$, as functions of the post-selected angle $\phi$, together with their difference $|\lambda_{+}| - |\lambda_{-}|$ and the imaginary part of the weak value $\mathrm{Im}(O_{s,w})$. In this plot $\alpha=\frac{\pi}{3}$, $\theta=\frac{\pi}{4}$ and $\gamma=0.1$, while the initial meter state is along the $z$ axis.}
    \label{fig:eigenvalues_and_imaginary_part_weak_value}
\end{figure}
As predicted analytically, the eigenvalues have equal absolute values whenever the imaginary part of the weak value vanishes (Fig.~\ref{fig:eigenvalues_and_imaginary_part_weak_value}). The sum of the weak value weighted by the post-selection probability over all post-selected states corresponds to the expectation value of the operator. However, the sum of the imaginary part over all polar angles $\phi$ is nonzero, because one must also sum over all azimuthal angles $\alpha$ to account for all possible post-selected states.

Next, the discussion turns to the dynamics of the quantum state on the Bloch sphere, aiming to clarify how its trajectory relates to the observed discontinuous behavior. The trajectory of the state on the Bloch sphere can be expressed as
\begin{eqnarray}
&&r_{x,n}=\text{Tr}\left[\hat{\sigma}_x\hat{\rho}_A^n\right]=\frac{\kappa r_{x,n-1}+\beta+\zeta}{\kappa+\left(\beta+\zeta\right)r_{x,n-1}} \\ \nonumber
&&r_{y,n}=\text{Tr}\left[\hat{\sigma}_y\hat{\rho}_A^n\right]=\frac{\kappa r_{y,n-1}+ir_{z,n-1}\left(\beta-\zeta\right)}{\kappa+\left(\beta+\zeta\right)r_{x,n-1}} \\ \nonumber
&&r_{z,n}=\text{Tr}\left[\hat{\sigma}_z\hat{\rho}_A^n\right]=\frac{\kappa r_{z,n-1}-ir_{y,n-1}\left(\beta-\zeta\right)}{\kappa+\left(\beta+\zeta\right)r_{x,n-1}}, 
\end{eqnarray}
where $\kappa=|c\left(\gamma\right)|^2+|d\left(\gamma\right)|^2$, $\beta=c\left(\gamma\right)d^{*}$, $\zeta=c^{*}\left(\gamma\right)d\left(\gamma\right)$.

\begin{figure}
    \centering
\includegraphics[width=0.9\linewidth]{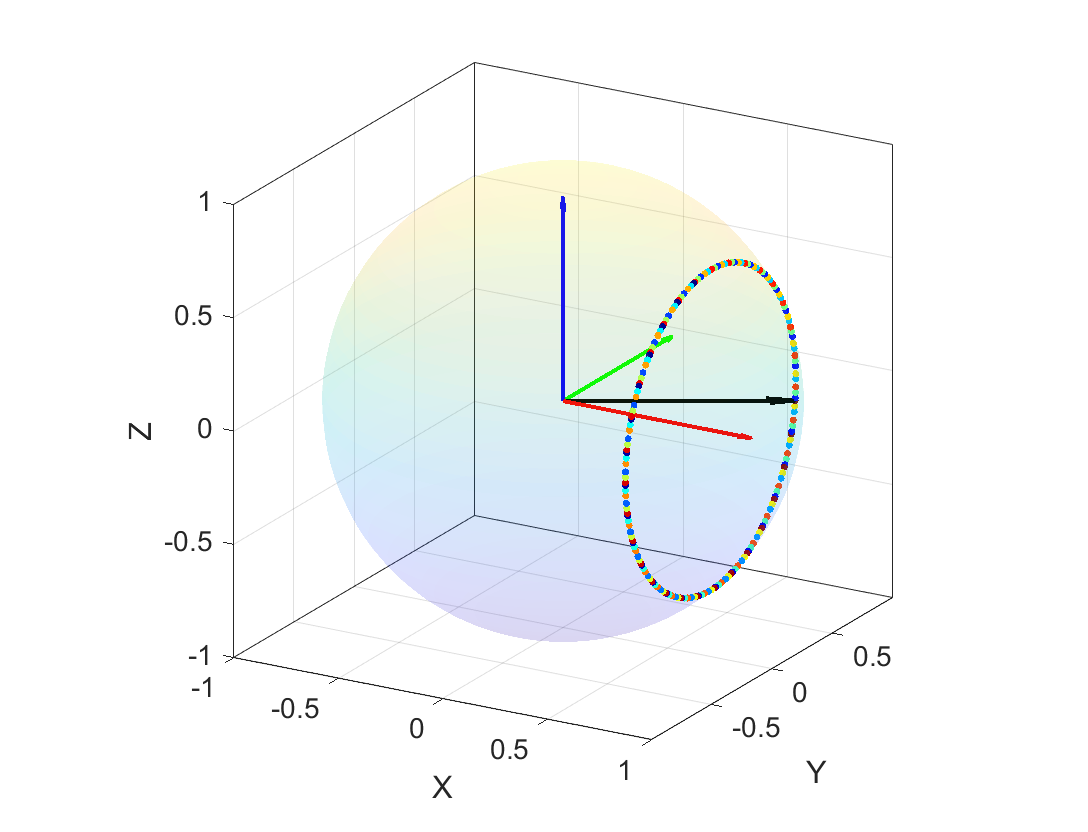}
    \caption{The Bloch sphere trajectory under unitary dynamics traces cycles at a constant polar angle, highlighting the periodic nature of the system. The path begins at the blue point and ends at the red point. System parameters are $\theta = \frac{\pi}{4}$, $\alpha = \frac{\pi}{7}$, $\gamma = 0.1$, $\phi = \frac{\pi}{2}$, with initial meter state $r = (\sqrt{0.5}, 0, \sqrt{0.5})^T$.}
    \label{fig:Bloch_sphere_closed_circle}
\end{figure}

Four distinct dynamical regimes emerge from the analysis. When the imaginary part of the weak value vanishes, the system exhibits purely unitary dynamics, resulting in periodic motion. In this case, the trajectories on the Bloch sphere are circular, with the initial parameter $\theta$ setting the starting point of the meter state (Fig.~\ref{fig:Bloch_sphere_closed_circle}). This regime occurs for $\phi = 0$, $\frac{\pi}{2}$, and $\pi$, where the eigenvalues of the operator $\hat K$ coincide. 

\begin{figure}
    \centering
\includegraphics[width=0.9\linewidth]{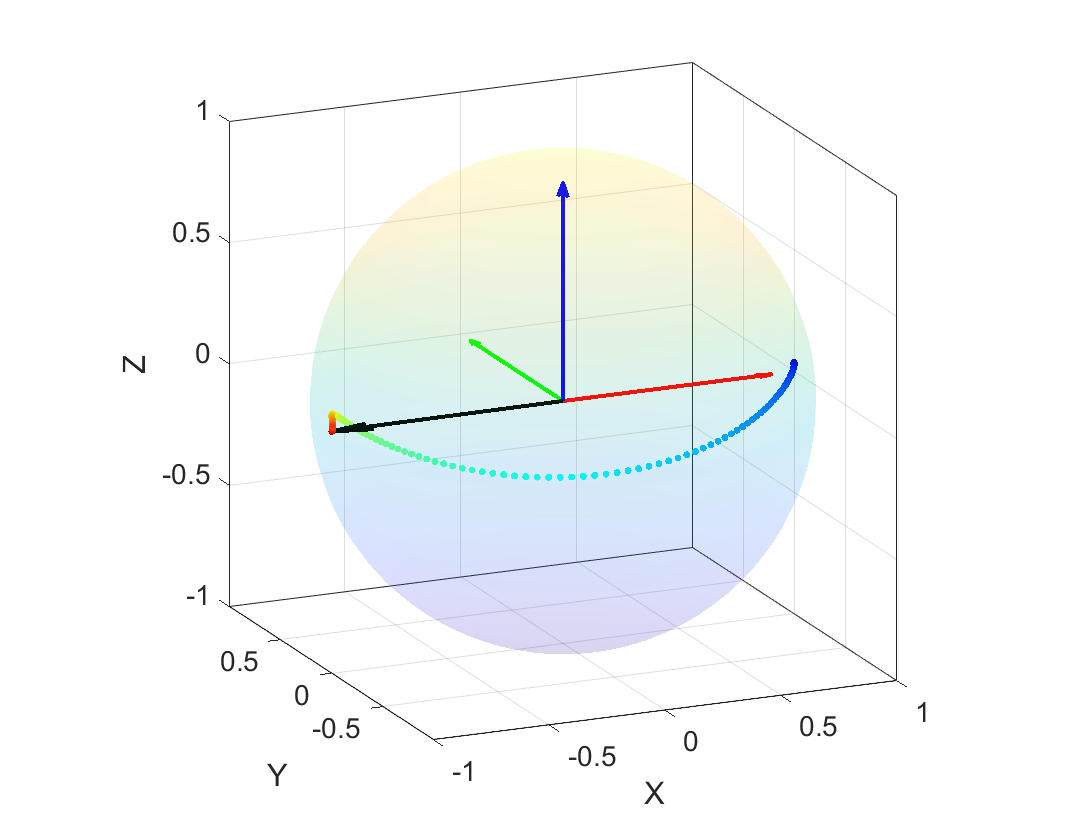}
    \caption{Although initiated near the unstable fixed point, the trajectory on the Bloch sphere is repelled toward the other fixed point, starting at the blue point and ending at the red point. Parameters: $\theta = \frac{\pi}{4}$, $\alpha = \frac{\pi}{7}$, $\gamma = 0.1$, $\phi = 1$, initial state $r = (\sqrt{0.99}, 0, \sqrt{0.001})^T$.}
    \label{fig:Bloch_trajectory_unstable_fixed_point}
\end{figure}

\begin{figure}
    \centering
\includegraphics[width=0.9\linewidth]{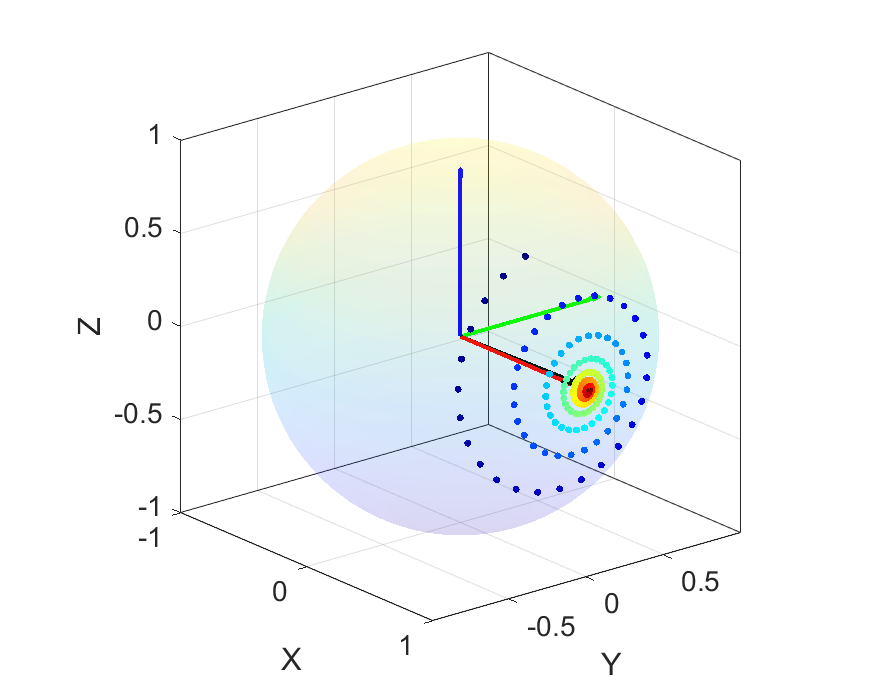}
    \caption{Starting near the unitary trajectory, the Bloch sphere path cycles while approaching the stable fixed point at $r = (1, 0, 0)^T$, beginning at the blue point and ending at the red point. Parameters: $\theta = \frac{\pi}{4}$, $\alpha = \frac{\pi}{7}$, $\gamma = 0.1$, $\phi = \frac{\pi}{2} + 0.1$, initial meter state $r = (\sqrt{0.5}, 0, \sqrt{0.5})^T$.}
    \label{fig:Bloch_sphere_closed_to_unitary}
\end{figure}

For values of $\phi$ such that $\phi < \frac{\pi}{2}$, two fixed points emerge, corresponding to the eigenvectors of $\hat{\sigma}_x$, the operator acting on the meter in Eq.~\ref{eq:unitary_operator}. The fixed point at $x=-1$ is stable, while the one at $x=1$ is unstable. Even if the state begins close to $x=1$, the dynamics will eventually drive it toward the stable fixed point at $x=-1$ (Fig.~\ref{fig:Bloch_trajectory_unstable_fixed_point}). The dynamics will remain at $x=1$ only when the system is initially exactly at that point. Conversely, for $\frac{\pi}{2} < \phi < \pi$, the stability of the fixed points is reversed: $x=1$ becomes stable and $x=-1$ unstable. After passing $\phi = \frac{\pi}{2}$, the stable and unstable fixed points effectively swap roles. Finally, in the near-unitary regime, when the dynamics are close to but not exactly unitary, the trajectories cycle on the Bloch sphere before eventually collapsing onto the stable fixed point (Fig.~\ref{fig:Bloch_sphere_closed_to_unitary}).

It should be emphasized that the eigenvectors of the Kraus operator $\hat K$ remain constant and identical for all parameter values. Consequently, the point at which the stability of the fixed points changes cannot be characterized as an exceptional point, since the eigenvectors do not vary at that point.

The plot shown in Fig.~\ref{fig:expectation_value_X} corresponds to the weak-measurement regime. However, as shown in Eq.~\ref{eq:Kraus_general_strength}, the dynamics can be conveniently analyzed for larger interaction strengths in this specific two-qubit case. For any value of $\gamma t$, the discontinuities persist in the three cases where the imaginary part of the weak value vanishes, namely $\phi = 0$, $\phi = \frac{\pi}{2}$, and $\phi = \pi$. Nevertheless, the detailed behavior of the expectation value after a few applications of the Kraus operator depends on the interaction strength. As $\gamma t$ increases, the curve may initially deviate further from the zero axis during the first iterations of the protocol. For even stronger interactions, however, the behavior approaches that of the weak-measurement limit again, reflecting the underlying periodic dependence on $\gamma t$.

\begin{figure}
    \centering
\includegraphics[width=0.9\linewidth]{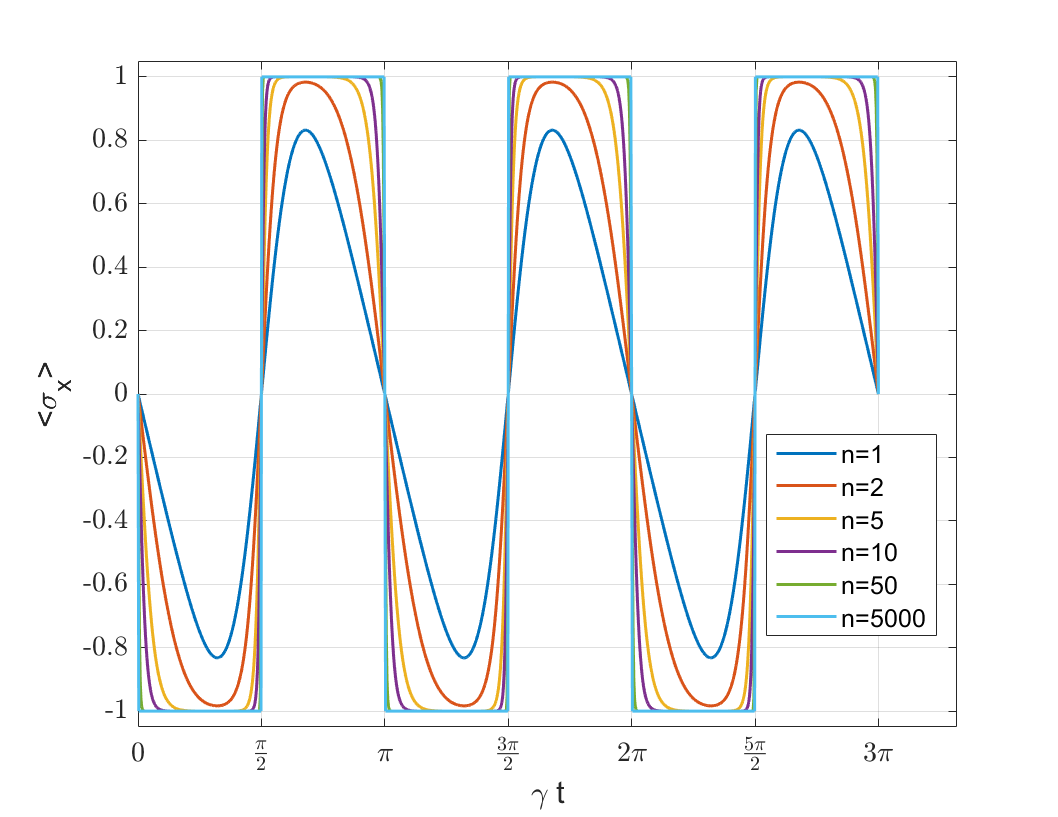}
    \caption{Discontinuities arise in the expectation value of $\hat{\sigma}_x$ as a function of $\gamma t$ at $\gamma t = m \frac{\pi}{2}$, where $m \in \mathbb{N}$, when the protocol is repeated a large number of times. In this plot, the parameters are fixed to $\alpha = \frac{\pi}{3}$, $\theta = \frac{\pi}{4}$, $\phi = \frac{\pi}{3}$, and the initial meter state is prepared along the z axis.}
    \label{fig:discontinuity_as_a_function_of_gamma}
\end{figure}

Moreover, the sign of the expectation value $\langle \hat{\sigma}_x \rangle$ in the intervals $\phi \in (0,\frac{\pi}{2})$ and $\phi \in (\frac{\pi}{2},\pi)$ changes at $\gamma t = \frac{\pi}{2}$, and subsequently alternates periodically with period $\frac{\pi}{2}$. At this point ($\gamma t = \frac{\pi}{2}$), the eigenvalue with the largest modulus changes, causing the previously unstable fixed point to become stable and vice versa. This exchange of stability gives rise to an additional discontinuity when $\gamma t$ is taken as the control parameter instead of $\phi$ (as shown in Fig.~\ref{fig:discontinuity_as_a_function_of_gamma}). For stronger interactions, the stability of the fixed points no longer depends solely on the sign of the imaginary part; rather, it becomes explicitly dependent on the interaction strength, leading to qualitative differences in stability across regimes. Fig.~\ref{fig:discontinuity_as_a_function_of_gamma} shows the expectation value as a function of $\gamma t$ for a representative point within the first window of Fig.~\ref{fig:expectation_value_X} $\left(\phi=\frac{\pi}{3}\right)$. As can be seen, discontinuities occur at intervals of $\frac{\pi}{2}$ in $\gamma t$. In contrast to the discontinuities observed when $\phi$ is used as the control parameter (Fig.~\ref{fig:expectation_value_X}), those appearing as a function of $\gamma t$ exhibit central symmetry (Fig.~\ref{fig:discontinuity_as_a_function_of_gamma}).
\section{Analysis of the Discontinuity}\label{section:analysis_discontinuity}
In second-order phase transitions, such as those in the Ising model~\cite{niss2005history} or the superconducting transition~\cite{kiometzis1994critical}, the behavior of observables near the critical point is characterized by critical exponents that are universal across systems belonging to the same universality class. Typically, near the critical point of a control parameter \(P\), for instance, the temperature, the magnetic field, or the interaction strength in measurement-induced phase transitions~\cite{poboiko2024measurement}, a characteristic physical quantity \(F\) (such as the heat capacity or the magnetic susceptibility) diverges following a power law in the vicinity of the critical point:
\begin{equation}
F\left(P\right) \propto |P - P_c|^{-\alpha},
\end{equation}
where \(P_c\) is the critical value of the control parameter and \(\alpha\) is the associated critical exponent.

In the protocol studied in this article, the post-selected polar angle \(\phi\) acts as the control parameter, governing a discontinuity in the expectation value of the Pauli matrix \(\hat{\sigma}_x\) at the critical point \(\phi_c\). This discontinuity emerges after multiple iterations of the protocol. The corresponding relaxation time, defined as the time required for the system to reach its steady state, diverges at \(\phi_c\), since at the critical point the meter dynamics become effectively unitary and no single eigenstate (fixed point) dominates. Although the discontinuity observed in Fig.~\ref{fig:expectation_value_X} might at first appear reminiscent of a first-order phase transition, the absence of a direct jump between two stable fixed points, and the emergence of unitary dynamics precisely at the critical point, indicate a resemblance to a second-order phase transition. The divergence of the relaxation time~\cite{dziarmaga2010dynamics}, demonstrated below, signifies critical slowing down, a defining feature of second-order phase transitions. Thus, the relaxation time $\tau$ serves as a key quantity for characterizing the critical behavior and determining the associated critical exponent, as well as for probing the universality of the transition.
This section demonstrates that the relaxation time associated with the proposed protocol universally follows the power-law behavior expressed in:
\begin{equation}
\label{eq:relaxation_time_exponent}
\tau(\phi) \propto |\phi - \phi_c|^{-\nu},
\end{equation}
where \(\nu\) is the critical exponent.

The precise definition of the relaxation time depends on the type of dynamics. Here, the dynamics proceed in discrete time steps, so \(\tau\) can be defined in terms of the spectrum of the operator driving the dynamics, specifically the Kraus operator. This analysis determines the critical exponent from the relaxation dynamics, in analogy with dynamic Markovian processes~\cite{prinz2011markov, levin2017markov}, where the key quantity is the ratio between the two leading eigenvalues of the Kraus operator. To quantify the convergence rate of the iterated Kraus map toward its fixed point, the relaxation time is defined in terms of the map’s spectral properties. Consider the unnormalized meter state after \(n\) applications of the Kraus operator:
\begin{equation}
\rho_n^{(\mathrm{un})} = \hat K^n \rho_0 \hat K^{\dagger n} = \sum_{i,j} \lambda_i^n \lambda_j^{*\,n} d^0_{ij} |\phi_i\rangle\langle \phi_j|,
\end{equation}
where \(\lambda_i\) and \(|\phi_i\rangle\) are the eigenvalues and eigenvectors of \(\hat K\), and \(d^0_{ij}\) are the coefficients of the initial meter density operator \(\rho_0\).

Denoting by \(|\lambda_1|\) and \(|\lambda_2|\) the eigenvalues of the Kraus operator $\hat K$ with the largest moduli, such that \(|\lambda_1| > |\lambda_2|\), the dominant diagonal term scales as \(|\lambda_1|^{2n}\), while the next-to-leading term scales as  \(|\lambda_2|^{2n}\). Consequently, the ratio between subdominant and dominant contributions decays as \((|\lambda_2|/|\lambda_1|)^{2n}\). Matching this discrete decay to an effective exponential relaxation \(e^{-n/\tau}\) defines the relaxation time \(\tau\):
\begin{equation}
\label{eq:relaxation_time}
\Big(\frac{|\lambda_2|^2}{|\lambda_1|^2}\Big)^n = e^{-n/\tau}
\quad\Longrightarrow\quad
\tau = \frac{1}{\log \!\left(\frac{|\lambda_1|^2}{|\lambda_2|^2}\right)}
     = \frac{1}{|\log R|},
\end{equation}
where \(R = \frac{|\lambda_1|^2}{|\lambda_2|^2}\), and \(\log\) denotes the natural logarithm.

Near each discontinuity, when both moduli of the eigenvalues are equal, the relaxation time exhibits a divergence typical of dynamic critical phenomena. The study investigates whether this divergence follows a power-law behavior, as expressed in Eq.~\ref{eq:relaxation_time_exponent}, considering the behavior both analytically, for any system governed by the protocol outlined in this article, and numerically, for the two-level system discussed in section~\ref{section:two_level_system}.

For any system evolving under the dynamics governed by the weak measurement protocol, the eigenvalues of the Kraus operator are given by Eq.~\ref{eq:general_eigenvalues_Kraus_operator}. We assume that the meter operator $\hat{O}_A$ is non-degenerate. In this case, let us denote the largest eigenvalue in modulus by $\lambda_1$ and the second-largest by $\lambda_2$, corresponding to the eigenvalues $o_1$ and $o_2$ of the operator $\hat{O}_A$, respectively. The squared moduli of these eigenvalues can then be expressed, to first order in $gt$, as
\begin{equation}
|\lambda_{1,2}|^2\approx |\braket{\psi_f|\psi_S}|^2\left(1+2gto_{1,2}\text{Im}\left(o_{S,w}\right)\right).
\end{equation}
Consequently, the function $R$ takes the form
\begin{equation}
R\approx \frac{\left(1+2gto_{1}\text{Im}\left(o_{S,w}\right)\right)}{\left(1+2gto_{2}\text{Im}\left(o_{S,w}\right)\right)}.
\end{equation}
To calculate the critical exponent near the point where the imaginary part of the weak value vanishes, one can expand $R$ in a Taylor series, assuming that $2 g t o_{2} \, \text{Im}\left(o_{S,w}\right)$ and $2 g t o_{1} \, \text{Im}\left(o_{S,w}\right)$ are small, and keeping terms to first order in $g t$ as
\begin{eqnarray}
R&\approx& \left(1+2gto_{1}\text{Im}\left(o_{S,w}\right)\right)\left(1-2gto_{2}\text{Im}\left(o_{S,w}\right)\right)\\ \nonumber
& \approx& 1 + 2gt\text{Im}\left(o_{S,w}\right)\left(o_{1}-o_{2}\right).
\end{eqnarray}
With this, the natural logarithm of $R$ can be further expanded in a Taylor series as
\begin{equation}
\log\left(R\right)\approx 2gt\text{Im}\left(o_{S,w}\right)\left(o_{1}-o_{2}\right),
\end{equation}
and, consequently, the relaxation time is given, to first order in $gt$, by
\begin{equation}
\tau\approx\frac{1}{|2gt\text{Im}\left(o_{S,w}\right)\left(o_{1}-o_{2}\right)|}.
\end{equation}
At $\phi = \phi_c$, the imaginary part of the weak value $\text{Im}\left(o_{S,w}\right)$, vanishes exactly. However, in typical cases of weak measurements, the weak value varies smoothly with the post-selected angle, so it is reasonable to assume that near $\phi_c$, the imaginary part of the weak value depends linearly on $\phi$ as
\begin{equation}
\text{Im}\left(o_{S,w}\right)\approx A|\phi-\phi_c|, 
\end{equation}
where $A$ is a real constant. In this case, the relaxation time depends on $\phi$ according to
\begin{equation}
\tau\approx\frac{1}{|2gtA\left(\phi-\phi_c\right)\left(o_{1}-o_{2}\right)|}\propto \frac{1}{|\phi-\phi_c|},
\end{equation}
which has the form of a power law, and by comparison with Eq.~\ref{eq:relaxation_time_exponent}, one can deduce that the critical exponent is $\nu = 1$, independent of the meter dimension and other system parameters, provided the assumptions made in this section are satisfied.

\begin{figure}[h!]
    \centering
    \includegraphics[width=0.9\linewidth]{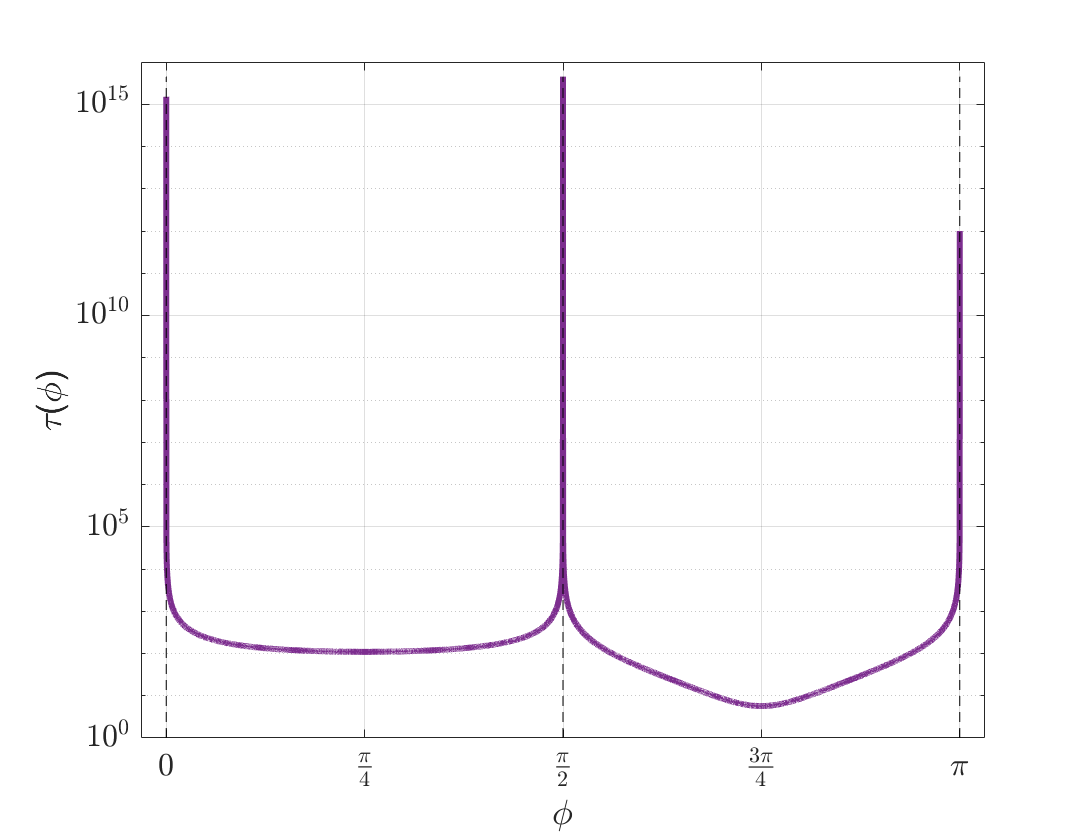}
    \caption{The relaxation time \(\tau\) exhibits three divergences as a function of the polar post-selection angle \(\phi\) at the critical points. The parameters used are \(\alpha = \pi/7\), \(\theta = \pi/4\), \(\gamma = 0.01\).}
    \label{fig:relaxation_time}
\end{figure}

To conclude this section, the qubit–qubit example of section~\ref{section:two_level_system} serves to illustrate this behavior. Three divergences appear in the relaxation time \(\tau\) as a function of the polar post-selection angle \(\phi\) (Fig.~\ref{fig:relaxation_time}). They correspond to the points where \(\mathrm{Im}(O_{S,w}) = 0\) in the two-level system studied in section~\ref{section:two_level_system}. At these points, the imaginary part of the weak value vanishes and the meter evolution becomes effectively unitary at first order in $gt$, coinciding with the discontinuities observed in the system’s expectation values.

\begin{figure}[h!]
    \centering
    \includegraphics[width=0.9\linewidth]{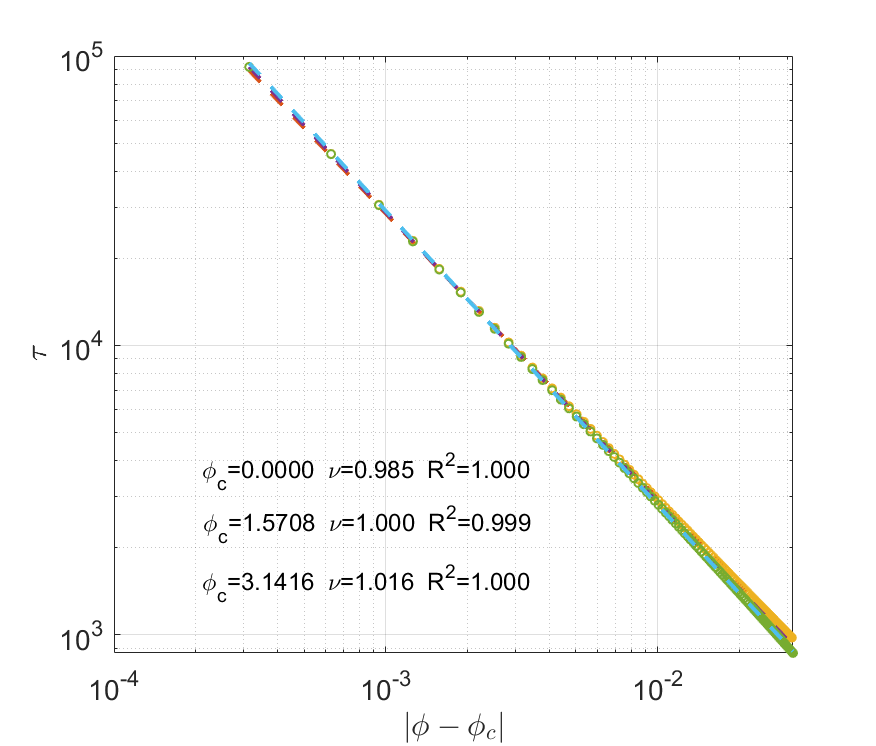}
    \caption{The relaxation time $\tau$ exhibits power-law scaling with exponent $\nu = 1$ near the three discontinuities, as shown in this log–log plot with linear regression fits. Parameters: $\alpha = \pi/7$, $\theta = \pi/4$, $\gamma = 0.01$.}
    \label{fig:linear_regression_relaxation_time}
\end{figure}

The relaxation time \(\tau\) near the three previously identified discontinuity points for the two-level system exhibits the expected power-law scaling, with linear fits giving a critical exponent of \(\nu = 1\) (Fig.~\ref{fig:linear_regression_relaxation_time}).
\section{Conclusion}\label{section:conclusion}
Complex weak values are essential for the emergence of dynamical discontinuities, whereas purely real weak values suppress them. The relaxation time diverges at the critical point and exhibits a scaling law that appears to be universal, independent of the protocol parameters, with a critical exponent of 1. This behavior is reminiscent of second order phase transitions, as has been demonstrated analytically for systems of any dimension and confirmed numerically.

A nonzero imaginary part of the weak value modifies the system's evolution from a process that is unitary at first order in $gt$, where the Kraus operator has real eigenvalues of equal modulus at first order in $gt$, to a non-unitary dynamics featuring $N$ fixed points, with $N$ denoting the dimension of the system's Hilbert space. Among these, one fixed point is stable, provided that the Kraus operator is non-degenerate. In this case, the dynamics drives the system toward the stable fixed point, which corresponds to the eigenvector associated with the largest eigenvalue of the Kraus operator, unless the meter is initially prepared in exactly one of the other eigenvectors corresponding to unstable fixed points. This eigenvector coincides with the smallest eigenvalue of the meter operator if the imaginary part of the weak value is positive, and with the largest eigenvalue if the imaginary part of the weak value is negative.

Extending the detailed analysis to larger meter dimensions could reveal connections to genuine many-body phase transitions in the thermodynamic limit.

The results demonstrate that the weak value can act as a tunable parameter capable of modifying the stability structure of measurement-induced quantum dynamics. By continuously adjusting the post-selection angle, the protocol induces abrupt changes in long-time meter observables, enabling controlled manipulation of the meter state and measurement-based switching within finite-dimensional quantum systems, without requiring many-body interactions or nonlinear Hamiltonians. The asymptotic states correspond to fixed points of the associated Kraus map, providing a reliable route for state preparation. Repeated post-selected measurements generate effective non-unitary dynamics that act as dissipation, and tuning the weak value determines which fixed point the system relaxes to, realizing a form of measurement-induced dissipative control. Near the critical point, the sharp dependence of observables on the post-selection angle enhances sensitivity to small variations, suggesting applications in high-precision parameter estimation. This protocol therefore provides a versatile platform for exploring critical-like phenomena and dynamical bifurcations in analytically tractable, few-body systems, and could potentially be realized in current experimental setups with small, well-controlled quantum systems.
\section*{Acknowledgments}
I thank Alexandre Matzkin for fruitful discussions that enriched this work. I am grateful to Sébastien R. Mouchet for a careful reading of the manuscript and providing valuable suggestions. This research was supported by the EUTOPIA Science and Innovation Fellowship Programme and funded by the European Union Horizon 2020 programme under the Marie Skłodowska-Curie grant agreement No 945380.
\appendix
\section{Discontinuities from eigenvalue exchange}\label{appendix:analysis_discontinuities}
Discontinuities in the long-time meter state occur precisely when the eigenvalue of $K(\phi)$ with the largest modulus ceases to be dominant and another eigenvalue takes its place.

Let $\hat K(\phi)$ denote a family of linear operators on the finite-dimensional Hilbert space of the meter, representing the Kraus operators determined by the parameters of the protocol. This family depends continuously on the real parameter $\phi$, the post-selected polar angle. The unnormalized map is
\begin{equation}
    \hat\rho_{A,n}^{\text{(un)}}(\phi) = \hat K(\phi)^n \hat\rho_{A,0} \hat K(\phi)^{\dagger n},
\end{equation}
for an arbitrary initial meter state $\hat\rho_{A,0}$. The normalized state after $n$ iterations is
\begin{equation}
    \hat\rho_{A,n}(\phi) = \frac{\hat\rho_{A,n}^{\text{(un)}}(\phi)}{\text{Tr}[\hat\rho_{A,n}^{\text{(un)}}(\phi)]}.
\end{equation}
Assume that, for each $\phi$, the operator $\hat K(\phi)$ is diagonalizable with eigenvalues $\{\lambda_j(\phi)\}$ and orthogonal eigenvectors $\ket{\lambda_j\left(\phi\right)}$, satisfying $\braket{\lambda_i(\phi)|\lambda_j(\phi)}=\delta_{ij}$. This assumption holds in the protocol, as the Kraus operators involve only the identity and an observable of the meter. This study focuses on non-degenerate observables.

Consider that, for $\phi<\phi_c$, the eigenvalue of $\hat K(\phi)$ with the largest modulus is unique and given by $\lambda_1(\phi)$, while for $\phi >\phi_c$ it is $\lambda_2(\phi)$, with
\begin{equation}
    |\lambda_1(\phi_c)| = |\lambda_2(\phi_c)|.
\end{equation}
Let the corresponding normalized projectors be
\begin{equation}
    \hat P^1 = \lim_{\phi\to\phi_c^-} \frac{|\lambda_1(\phi)\rangle\langle \lambda_1(\phi)|}{\langle \lambda_1(\phi)|\lambda_1(\phi)\rangle}
    \quad \neq \quad 
    \lim_{\phi\to\phi_c^+} \frac{|\lambda_2(\phi)\rangle\langle \lambda_2(\phi)|}{\langle \lambda_2(\phi)|\lambda_2(\phi)\rangle} = \hat P^2,
\end{equation}
and assume $\langle \lambda_j(\phi)|\hat\rho_{A,0}|\lambda_j(\phi)\rangle \neq 0$.

Since $\hat K(\phi)$ is diagonalizable,
\begin{equation}
    \hat K(\phi)^n = \sum_j \lambda_j(\phi)^n 
    |\lambda_j(\phi)\rangle\langle \lambda_j(\phi)|,
\end{equation}
and hence
\begin{equation}
    \hat\rho_{A,n}^{(\mathrm{un})}(\phi)
    = \sum_{i,j} 
    \lambda_i(\phi)^n \lambda_j(\phi)^{*n}
    c_{ij}(\phi)
    |\lambda_i(\phi)\rangle\langle \lambda_j(\phi)|,
\end{equation}
with $c_{ij}(\phi)=\langle \lambda_i(\phi)|\hat\rho_{A,0}|\lambda_j(\phi)\rangle$.

For $\phi<\phi_c$, the term with $|\lambda_1(\phi)|$ dominates as $n\to\infty$, leading to
\begin{equation}
    \hat\rho_{A,\infty}(\phi)
    = \frac{|\lambda_1(\phi)\rangle\langle \lambda_1(\phi)|}
           {\langle \lambda_1(\phi)|\lambda_1(\phi)\rangle},
\end{equation}
and, similarly for $\phi>\phi_c$, the dominant contribution comes from 
$\lambda_2(\phi)$.

Because the limiting projectors on either side of $\phi_c$ differ, the long-time meter state exhibits a discontinuity at $\phi_c$, which directly transfers to the expectation values of observables.
\bibliographystyle{unsrt}
\bibliography{biblio}

@article{aharonov1988result,
  title={How the result of a measurement of a component of the spin of a spin-1/2 particle can turn out to be 100},
  author={Aharonov, Yakir and Albert, David Z and Vaidman, Lev},
  journal={Physical review letters},
  volume={60},
  number={14},
  pages={1351},
  year={1988},
  publisher={APS}
}

@article{aharonov2002revisiting,
  title={Revisiting Hardy's paradox: counterfactual statements, real measurements, entanglement and weak values},
  author={Aharonov, Yakir and Botero, Alonso and Popescu, Sandu and Reznik, Benni and Tollaksen, Jeff},
  journal={Physics Letters A},
  volume={301},
  number={3-4},
  pages={130--138},
  year={2002},
  publisher={Elsevier}
}

@article{matzkin2012observing,
  title={Observing Trajectories with Weak Measurements in Quantum Systems in the Semiclassical Regime},
  author={Matzkin, A},
  journal={Physical review letters},
  volume={109},
  number={15},
  pages={150407},
  year={2012},
  publisher={APS}
}

@article{dressel2015weak,
  title={Weak values as interference phenomena},
  author={Dressel, Justin},
  journal={Physical Review A},
  volume={91},
  number={3},
  pages={032116},
  year={2015},
  publisher={APS}
}

@article{jordan2014technical,
  title={Technical advantages for weak-value amplification: when less is more},
  author={Jordan, Andrew N and Mart{\'\i}nez-Rinc{\'o}n, Juli{\'a}n and Howell, John C},
  journal={Physical Review X},
  volume={4},
  number={1},
  pages={011031},
  year={2014},
  publisher={APS}
}

@article{dixon2009ultrasensitive,
  title={Ultrasensitive beam deflection measurement via interferometric weak value amplification},
  author={Dixon, P Ben and Starling, David J and Jordan, Andrew N and Howell, John C},
  journal={Physical review letters},
  volume={102},
  number={17},
  pages={173601},
  year={2009},
  publisher={APS}
}

@article{luo2017precision,
  title={Precision improvement of surface plasmon resonance sensors based on weak-value amplification},
  author={Luo, Lan and Qiu, Xiaodong and Xie, Linguo and Liu, Xiong and Li, Zhaoxue and Zhang, Zhiyou and Du, Jinglei},
  journal={Optics Express},
  volume={25},
  number={18},
  pages={21107--21114},
  year={2017},
  publisher={Optical Society of America}
}

@article{pusey2014anomalous,
  title={Anomalous weak values are proofs of contextuality},
  author={Pusey, Matthew F},
  journal={Physical review letters},
  volume={113},
  number={20},
  pages={200401},
  year={2014},
  publisher={APS}
}

@article{kunjwal2019anomalous,
  title={Anomalous weak values and contextuality: Robustness, tightness, and imaginary parts},
  author={Kunjwal, Ravi and Lostaglio, Matteo and Pusey, Matthew F},
  journal={Physical Review A},
  volume={100},
  number={4},
  pages={042116},
  year={2019},
  publisher={APS}
}

@article{arvidsson2024properties,
  title={Properties and applications of the Kirkwood--Dirac distribution},
  author={Arvidsson-Shukur, David RM and Braasch Jr, William F and De Bievre, Stephan and Dressel, Justin and Jordan, Andrew N and Langrenez, Christopher and Lostaglio, Matteo and Lundeen, Jeff S and Halpern, Nicole Yunger},
  journal={New Journal of Physics},
  volume={26},
  number={12},
  pages={121201},
  year={2024},
  publisher={IOP Publishing}
}

@article{ferraz2022geometrical,
  title={Geometrical interpretation of the argument of weak values of general observables in $N$-level quantum systems},
  author={Ferraz, Lorena Ballesteros and Lambert, Dominique L and Caudano, Yves},
  journal={Quantum Science and Technology},
  volume={7},
  number={4},
  pages={045028},
  year={2022},
  publisher={IOP Publishing}
}

@article{ferraz2023revisiting,
  title={Revisiting weak values through non-normality},
  author={Ferraz, Lorena Ballesteros and Muolo, Riccardo and Caudano, Yves and Carletti, Timoteo},
  journal={Journal of Physics A: Mathematical and Theoretical},
  volume={56},
  number={47},
  pages={475303},
  year={2023},
  publisher={IOP Publishing}
}

@article{ferraz2025exploring,
  title={Exploring weak value arguments and Bargmann invariants in $N$-level quantum systems through the Majorana symmetric representation},
  author={Ferraz, Lorena Ballesteros and Lambert, Dominique and Caudano, Yves},
  journal={Journal of Physics A: Mathematical and Theoretical},
  volume={58},
  number={20},
  pages={205303},
  year={2025},
  publisher={IOP Publishing}
}

@article{ferraz2024relevance,
  title={On the relevance of weak measurements in dissipative quantum systems},
  author={Ferraz, Lorena Ballesteros and Martin, John and Caudano, Yves},
  journal={Quantum Science and Technology},
  volume={9},
  number={3},
  pages={035029},
  year={2024},
  publisher={IOP Publishing}
}

@article{heyl2013dynamical,
  title={Dynamical quantum phase transitions in the transverse-field Ising model},
  author={Heyl, Markus and Polkovnikov, Anatoli and Kehrein, Stefan},
  journal={Physical review letters},
  volume={110},
  number={13},
  pages={135704},
  year={2013},
  publisher={APS}
}

@article{zvyagin2016dynamical,
  title={Dynamical quantum phase transitions},
  author={Zvyagin, AA},
  journal={Low Temperature Physics},
  volume={42},
  number={11},
  pages={971--994},
  year={2016},
  publisher={AIP Publishing}
}

@article{jurcevic2017direct,
  title={Direct observation of dynamical quantum phase transitions in an interacting many-body system},
  author={Jurcevic, P and Shen, H and Hauke, P and Maier, C and Brydges, T and Hempel, C and Lanyon, BP and Heyl, Markus and Blatt, R and Roos, CF},
  journal={Physical review letters},
  volume={119},
  number={8},
  pages={080501},
  year={2017},
  publisher={APS}
}

@article{skinner2019measurement,
  title={Measurement-induced phase transitions in the dynamics of entanglement},
  author={Skinner, Brian and Ruhman, Jonathan and Nahum, Adam},
  journal={Physical Review X},
  volume={9},
  number={3},
  pages={031009},
  year={2019},
  publisher={APS}
}

@article{block2022measurement,
  title={Measurement-induced transition in long-range interacting quantum circuits},
  author={Block, Maxwell and Bao, Yimu and Choi, Soonwon and Altman, Ehud and Yao, Norman Y},
  journal={Physical Review Letters},
  volume={128},
  number={1},
  pages={010604},
  year={2022},
  publisher={APS}
}

@article{li2023cross,
  title={Cross entropy benchmark for measurement-induced phase transitions},
  author={Li, Yaodong and Zou, Yijian and Glorioso, Paolo and Altman, Ehud and Fisher, Matthew PA},
  journal={Physical Review Letters},
  volume={130},
  number={22},
  pages={220404},
  year={2023},
  publisher={APS}
}

@article{noel2022measurement,
  title={Measurement-induced quantum phases realized in a trapped-ion quantum computer},
  author={Noel, Crystal and Niroula, Pradeep and Zhu, Daiwei and Risinger, Andrew and Egan, Laird and Biswas, Debopriyo and Cetina, Marko and Gorshkov, Alexey V and Gullans, Michael J and Huse, David A and others},
  journal={Nature Physics},
  volume={18},
  number={7},
  pages={760--764},
  year={2022},
  publisher={Nature Publishing Group UK London}
}

@article{koh2023measurement,
  title={Measurement-induced entanglement phase transition on a superconducting quantum processor with mid-circuit readout},
  author={Koh, Jin Ming and Sun, Shi-Ning and Motta, Mario and Minnich, Austin J},
  journal={Nature Physics},
  volume={19},
  number={9},
  pages={1314--1319},
  year={2023},
  publisher={Nature Publishing Group UK London}
}

@article{puebla2020finite,
  title={Finite-component dynamical quantum phase transitions},
  author={Puebla, Ricardo},
  journal={Physical Review B},
  volume={102},
  number={22},
  pages={220302},
  year={2020},
  publisher={APS}
}

@article{prinz2011markov,
  title={Markov models of molecular kinetics: Generation and validation},
  author={Prinz, Jan-Hendrik and Wu, Hao and Sarich, Marco and Keller, Bettina and Senne, Martin and Held, Martin and Chodera, John D and Sch{\"u}tte, Christof and No{\'e}, Frank},
  journal={The Journal of chemical physics},
  volume={134},
  number={17},
  year={2011},
  publisher={AIP Publishing}
}

@book{levin2017markov,
  title={Markov chains and mixing times},
  author={Levin, David A and Peres, Yuval},
  volume={107},
  year={2017},
  publisher={American Mathematical Soc.}
}

@article{Bauer2011,
  title = {Convergence of repeated quantum nondemolition measurements and wave function collapse},
  author = {Bauer, Michel and Bernard, Denis},
  journal = {Phys. Rev. A},
  volume = {84},
  pages = {044103},
  year = {2011},
  doi = {10.1103/PhysRevA.84.044103}
}

@article{Bauer2012,
  title = {Repeated Quantum Non-Demolition Measurements: Convergence and Continuous Time Limit},
  author = {Bauer, Michel and Bernard, Denis},
  journal = {Annales Henri Poincaré},
  volume = {13},
  pages = {1799--1821},
  year = {2012},
  doi = {10.1007/s00023-012-0204-x}
}

@article{Szyniszewski2019,
  title = {Entanglement transition from variable-strength weak measurements},
  author = {Szyniszewski, M. and Romito, A. and Schomerus, H.},
  journal = {Phys. Rev. B},
  volume = {100},
  pages = {064204},
  year = {2019},
  doi = {10.1103/PhysRevB.100.064204}
}

@article{Dubey2021,
  title = {Quantum dynamics under continuous projective measurements: Non-Hermitian description and phase transitions},
  author = {Dubey, Vimal and Bernardin, C. and Dhar, Abhishek},
  journal = {Phys. Rev. A},
  volume = {103},
  pages = {032221},
  year = {2021},
  doi = {10.1103/PhysRevA.103.032221}
}

@article{Roux2024,
  title = {Partial Wavefunction Collapse Under Repeated Weak Measurement of a Non-Conserved Observable},
  author = {Roux, Guillaume and et al.},
  journal = {arXiv preprint},
  eprint = {2412.05226},
  year = {2024},
  url = {https://arxiv.org/abs/2412.05226}
}

@article{Hertz1976,
  title = {Quantum critical phenomena},
  author = {John A. Hertz},
  journal = {Phys. Rev. B},
  volume={14, 1165},
  pages={1165},
  year = {1976},
  url = {https://journals.aps.org/prb/abstract/10.1103/PhysRevB.14.1165}
}

@article{millis1993effect,
  title={Effect of a nonzero temperature on quantum critical points in itinerant fermion systems},
  author={Millis, AJ},
  journal={Physical Review B},
  volume={48},
  number={10},
  pages={7183},
  year={1993},
  publisher={APS}
}

@article{sachdev1999quantum,
  title={Quantum phase transitions},
  author={Sachdev, Subir},
  journal={Physics world},
  volume={12},
  number={4},
  pages={33},
  year={1999},
  publisher={IOP Publishing}
}

@article{vojta2003quantum,
  title={Quantum phase transitions},
  author={Vojta, Matthias},
  journal={Reports on Progress in Physics},
  volume={66},
  number={12},
  pages={2069},
  year={2003},
  publisher={IOP Publishing}
}

@article{vzunkovivc2016dynamical,
  title={Dynamical phase transitions and Loschmidt echo in the infinite-range XY model},
  author={{\v{Z}}unkovi{\v{c}}, Bojan and Silva, Alessandro and Fabrizio, Michele},
  journal={Philosophical Transactions of the Royal Society A: Mathematical, Physical and Engineering Sciences},
  volume={374},
  number={2069},
  pages={20150160},
  year={2016},
  publisher={The Royal Society Publishing}
}

@article{de2021entanglement,
  title={Entanglement view of dynamical quantum phase transitions},
  author={De Nicola, Stefano and Michailidis, Alexios A and Serbyn, Maksym},
  journal={Physical Review Letters},
  volume={126},
  number={4},
  pages={040602},
  year={2021},
  publisher={APS}
}

@article{kawabata2023entanglement,
  title={Entanglement phase transition induced by the non-Hermitian skin effect},
  author={Kawabata, Kohei and Numasawa, Tokiro and Ryu, Shinsei},
  journal={Physical Review X},
  volume={13},
  number={2},
  pages={021007},
  year={2023},
  publisher={APS}
}

@article{niss2005history,
  title={History of the Lenz-Ising model 1920--1950: from ferromagnetic to cooperative phenomena},
  author={Niss, Martin},
  journal={Archive for history of exact sciences},
  volume={59},
  number={3},
  pages={267--318},
  year={2005},
  publisher={Springer}
}

@article{kiometzis1994critical,
  title={Critical exponents of the superconducting phase transition},
  author={Kiometzis, Michael and Kleinert, Hagen and Schakel, Adriaan MJ},
  journal={Physical review letters},
  volume={73},
  number={14},
  pages={1975},
  year={1994},
  publisher={APS}
}

@article{poboiko2024measurement,
  title={Measurement-induced phase transition for free fermions above one dimension},
  author={Poboiko, Igor and Gornyi, Igor V and Mirlin, Alexander D},
  journal={Physical Review Letters},
  volume={132},
  number={11},
  pages={110403},
  year={2024},
  publisher={APS}
}

@article{dziarmaga2010dynamics,
  title={Dynamics of a quantum phase transition and relaxation to a steady state},
  author={Dziarmaga, Jacek},
  journal={Advances in Physics},
  volume={59},
  number={6},
  pages={1063--1189},
  year={2010},
  publisher={Taylor \& Francis}
}

@article{hamazaki2021exceptional,
  title={Exceptional dynamical quantum phase transitions in periodically driven systems},
  author={Hamazaki, Ryusuke},
  journal={Nature communications},
  volume={12},
  number={1},
  pages={5108},
  year={2021},
  publisher={Nature Publishing Group UK London}
}

@article{di2024environment,
  title={Environment induced dynamical quantum phase transitions in two-qubit Rabi model},
  author={Di Bello, Grazia and Ponticelli, Andrea and Pavan, Fabrizio and Cataudella, Vittorio and De Filippis, Giulio and de Candia, Antonio and Perroni, Carmine Antonio},
  journal={Communications Physics},
  volume={7},
  number={1},
  pages={364},
  year={2024},
  publisher={Nature Publishing Group UK London}
}

@article{jozsa2007complex,
  title={Complex weak values in quantum measurement},
  author={Jozsa, Richard},
  journal={Physical Review A—Atomic, Molecular, and Optical Physics},
  volume={76},
  number={4},
  pages={044103},
  year={2007},
  publisher={APS}
}

@article{hofmann2011role,
  title={On the role of complex phases in the quantum statistics of weak measurements},
  author={Hofmann, Holger F},
  journal={New Journal of Physics},
  volume={13},
  number={10},
  pages={103009},
  year={2011},
  publisher={IOP Publishing}
}

@article{hofmann2012complex,
  title={Complex joint probabilities as expressions of reversible transformations in quantum mechanics},
  author={Hofmann, Holger F},
  journal={New Journal of Physics},
  volume={14},
  number={4},
  pages={043031},
  year={2012},
  publisher={IOP Publishing}
}

@article{hance2024counterfactuality,
  title={Counterfactuality, back-action, and information gain in multi-path interferometers},
  author={Hance, Jonte R and Matsushita, Tomonori and Hofmann, Holger F},
  journal={Quantum Science and Technology},
  volume={9},
  number={4},
  pages={045015},
  year={2024},
  publisher={IOP Publishing}
}

@article{hance2023contextuality,
  title={Contextuality, coherences, and quantum cheshire cats},
  author={Hance, Jonte R and Ji, Ming and Hofmann, Holger F},
  journal={New Journal of Physics},
  volume={25},
  number={11},
  pages={113028},
  year={2023},
  publisher={IOP Publishing}
}

@article{kofman2012nonperturbative,
  title={Nonperturbative theory of weak pre-and post-selected measurements},
  author={Kofman, Abraham G and Ashhab, Sahel and Nori, Franco},
  journal={Physics Reports},
  volume={520},
  number={2},
  pages={43--133},
  year={2012},
  publisher={Elsevier}
}
\end{document}